\documentclass[a4paper,twoside]{article}

\usepackage{graphicx}
\usepackage{hhline,theorem,verbatim,euscript,url}
\usepackage[babel=true]{csquotes}
\usepackage{listings}
\usepackage{amssymb}
\usepackage{amstext}
\usepackage{mathtools}
\usepackage{textcomp}
\usepackage[ruled,vlined,linesnumbered]{algorithm2e}
\SetAlFnt{\footnotesize\sf}
\usepackage{caption}
\usepackage{float} 
\usepackage{color}
\usepackage{textcomp}
\usepackage{fancyvrb}
\usepackage{array, multirow,tabularx,csquotes,makecell}
\usepackage[tight]{subfigure}
\usepackage{epsfig}
\usepackage{xtab,afterpage}

\usepackage{calc}

\usepackage{multicol}

\usepackage{pslatex}
\usepackage{apalike}
\newcolumntype{P}[1]{>{\raggedright\arraybackslash}p{#1}}
\newtheorem{theorem}{Theorem}
\newtheorem{definition}[theorem]{Definition}

\usepackage{SCITEPRESS}

\makeatletter
\newcommand{\xnrightarrow}[2][]{%
	\mathrel{%
		\vphantom{\xrightarrow[#1]{#2}}%
		\ooalign{\hidewidth\neg@arrow\hidewidth\cr$\m@th\xrightarrow[#1]{#2}$\cr}%
	}%
}
\newcommand{\neg@arrow}{%
	$\m@th\vcenter{\hbox{%
			\rotatebox[origin=c]{-45}{\scalebox{1.5}[1]{$\m@th\scriptscriptstyle|$}}%
	}}$
}
\makeatother

\begin{document}
	\title{Security Testing of RESTful APIs With Test Case Mutation}
\author{\authorname{S\'ebastien Salva\sup{1} and Jarod Sue\sup{1}}
	\affiliation{\sup{1}LIMOS - UMR CNRS 6158, Clermont Auvergne University, UCA, France}
\affiliation{\sup{2}Department of Computing, Main University, MySecondTown, MyCountry}
	\email{sebastien.salva@uca.fr, jarod.sue@uca.fr}
}

\abstract{
The focus of this paper is on automating the security testing of RESTful APIs. The testing stage of this specific kind of components is often performed manually, and this is yet considered as a long and difficult activity. This paper proposes an automated approach to help developers generate test cases for experimenting with each service in isolation. This approach is based upon the notion of test case mutation, which automatically generates new test cases from an original test case set. Test case mutation operators perform slight test case modifications to mimic possible failures or to test the component under test with new interactions. In this paper, we examine test case mutation operators for RESTful APIs and define 17 operators specialised in security testing. Then, we present our test case mutation algorithm. We evaluate its effectiveness and performance on four web service compositions.}
\keywords{RESTful APIs; Security; Test Case Generation; Test Case Mutation}

\onecolumn \maketitle \normalsize \vfill

\section{\uppercase{Introduction}}

One of the key motivations for software security is the prevention of attackers exploiting software flaws, which can lead to compromising application security or revealing user data. Despite the continuous growth of the security testing market, there is still an inadequate emphasis on this activity, exposing organisations and end users to unforeseen risks when using vulnerable systems or software. One aspect that may account for this observation, 
is that selecting security solutions and crafting specific security test cases are two tasks of the software life cycle that demand time, expertise, and experience. Developers often lack the guidance, resources, or skills on how to design, implement secure applications, and test them. Furthermore, different kinds of expertise are required, e.g., to represent threats, to choose the most appropriate security solutions w.r.t. an application context, or to ensure that the latter are implemented as expected.

A way to help developers in security testing is the use of test automation, which addresses challenges related to time constraints, complexity, and coverage. Model based testing \cite{mbt} offers the advantage of automating the test case generation. But models are often manually written, and this task is considered as long, difficult and error-prone, even for experts. Instead, fuzzing and automated penetration testing approaches do not require models. They provide random or malformed data as input or simulate attacks to assess the application or system security. Despite their significant benefits, a recurring limitation observed in employing these approaches is the insufficient understanding of the application business logic and context. As a result, they may fail to identify certain security vulnerabilities that require a deeper understanding of how the application behaves.

Focusing on this background, we propose an intermediate solution based upon the notion of test case mutation. Unlike mutation testing that aims at evaluating the effectiveness of an existing test case set by introducing intentional errors into the original source code of an application under test \cite{papadakis2019mutation}, test case mutation automatically generates new test cases from an original test case set. As the original  test cases should encode some knowledge about the application under test, the mutated test cases should deeper cover the application behaviours and features and hence should detect further defects. A test case mutation operator performs slight test case modifications to mimic possible failures or to experiment the system under test with new interactions. Some test case mutation based approaches have been proposed for detecting bugs or crashes \cite{10.1145/2786805.2803206,10.1145/1882291.1882330,ARCURI2018195,10.1145/3293455,FASE2018,Paiva20}. None of them deals with security testing.

In this paper, we propose a new approach, specifically designed for testing the security of RESTful APIs in isolation. This firstly implies that we propose new specific mutation operators devoted to detecting security issues or weaknesses. This also means that our approach generates new executable test cases but also \textit{mock components}. We recall that a mock component aims at simulating an existing component, while behaving in a predefined and controlled way to make testing more effective and efficient. Mocks are often used by developers to make test development easier or to increase test coverage. They may indeed be used to simplify the dependencies that make testing difficult (e.g., infrastructure or environment related dependencies). Besides, mocks are used to increase test efficiency by replacing slow-to-access components. In summary, the main contributions of this paper include:

\begin{enumerate}
	\item a study on mutation operators specialised in the security testing of RESTful APIs,  including the definition of 17 operators,
	\item an algorithm for the generation of mutated test cases along with test scripts and mock components,
	\item the implementation of the approach, along with 4 RESTful API compositions and Log files publicly available in \cite{companion},
	\item an evaluation with these 4 compositions (15 services) of its effectiveness (amount of generated mutated test cases, ability to uncover new security weaknesses or to further cover the service codes) and its performance.
\end{enumerate}

The paper is organised as follows:  we discuss the related work in Section \ref{sec:relatedwork}. We study and propose test case mutation operators for RESTful APIs in Section~\ref{sec:WSmutation}. Our test case mutation algorithm is presented in Section~\ref{sec:approach}. Section~\ref{sec:evaluation} presents our evaluation. Section~\ref{sec:conclusion} summarises our contributions and draws some perspectives for future work.

\section{\uppercase{Related work}}
	\label{sec:relatedwork}
	
Numerous approaches have been proposed to generate test cases without specification, for example by using random testing \cite{arcuri2011random}, model learning \cite{PetrenkoA19}, graphical user interfaces exploration \cite{salva:hal-02019705,FerreiraP19}, or test case mutation, which is the topic of this paper. As stated in the introduction, test case mutation should not be confused with mutation testing \cite{papadakis2019mutation,7899041}. The former approach takes as input an existing test case set and applies mutation operators to derive new mutated test cases, a.k.a. mutants, mostly used for robustness (crash detection) or performance testing. The later mutes implementations with other kinds of operators to evaluate the quality of a test case set. This paper proposes an approach that belongs to the first category. 

Two testing perspectives are considered in the test case mutation approaches available in the literature, which offers a simple way to classify them. 

Some approaches consider white box testing \cite{10.1145/2786805.2803206,10.1145/1882291.1882330,ARCURI2018195,10.1145/3293455}. Having access to the source code indeed offers the strong advantage to being able to evaluate the relevance of the mutants by measuring code coverage. In \cite{10.1145/2786805.2803206}, a test case set is derived from stack traces by keeping only the test cases that experiment a given class. Test cases are then mutated by means of 5 operators to produce further tests specialised in crash testing. \cite{10.1145/1882291.1882330} compare two test case augmentation methods, one using concolic testing and another one using genetic algorithms that generate mutants. 
The evaluation of both approaches shows that the use of mutants is more effective to detect new bugs. \cite{ARCURI2018195} proposed algorithms to create test suites for Web services by considering the test case generation as a  multi-objective problem, whose objectives are related to metrics over source code properties (branch coverage, time limit). 
EVOMaster \cite{10.1145/3293455} implements this algorithm to generate robustness tests for RESTful APIs. 

All of these approaches require the source code, which is not always available. Hence, other approaches are based on black box testing, which is the case for our algorithm. The approach proposed by \cite{FASE2018} mutates existing test cases for mobile applications  with 6 operators for Android systems. The new tests aim at uncovering unexpected crashes, e.g., unhandled exceptions or network-based crashes.
\cite{Paiva20} proposed to mutate test cases for checking Web sites are robust to unexpected events. Generic test cases are firstly extracted from existing user executions. These are converted into concrete test cases by using test data generators. These test cases are then mutated to get new test cases that mimic specific problems, e.g. wrong passwords, removal of a request, etc. 

Another body of related work addresses the mutation of models \cite{TestingSoftwareModellingToolsUsingDataMutation,SITV17}, from which test cases can be later generated, for example with model based testing. But writing accurate and comprehensive models that represent the behaviour of a system is often long and complex.

Surprisingly, we did not find any test case mutation approach dedicated to security testing. Furthermore, none of the previous approaches consider mock components, yet these are massively used in the Industry with isolation testing. We hence contribute in this topic by firstly studying and proposing a list of mutation operators dedicated to testing the security of RESTful APIs. Secondly, we propose an algorithm to generate new test cases along with new mock components by means of mutation operators. We also propose strategies to limit the number of the generated mutants. 




\section{\uppercase{Test Case Mutation Operators for RESTful APIs}}
\label{sec:WSmutation}



This paper focuses on test case mutation operators designed to detect security weaknesses in RESTful APIs. This testing context introduces specific requirements and a testing architecture, both of which are subsequently presented. From this architecture, we present how test cases are modelled with Input Ouput Transition Systems (IOTSs) and provide an illustrative example. Then, we study the mutation operators that can be defined within this scope.

\subsection{Assumptions}

\begin{figure}[htbp]
\vspace{-0.4cm}
	\begin{center}
		\includegraphics[width=.7\linewidth]{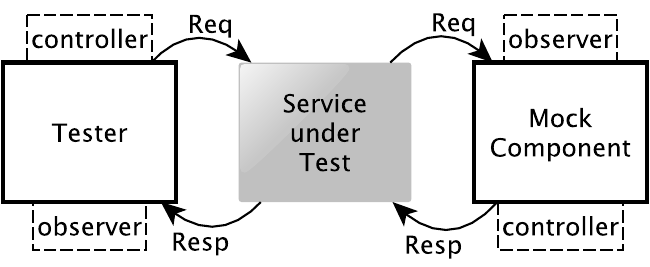}
	\end{center}
	\caption{Black-box test architecture for experimenting RESTful API in isolation}
	\label{fig:archi}
\end{figure}

We consider the test architecture depicted in Figure \ref{fig:archi}, whose attributes are expressed with the following realistic assumptions:

\begin{itemize}
	\item \textbf{Black box testing:} we employ a black box perspective, enabling to interact with a RESTful API, denoted $SUT$, only with HTTP requests or responses. We call them (communication) events;
	
	\item \textbf{Event content:} observers are able to get all the events related to a RESTful API under test along with their contents (no encryption). In particular, events include parameter assignments allowing to identify the source and the destination of each event. Besides, an event can be identified either as a request or a response;
	
	\item \textbf{Test in isolation:} we consider conducting tests in an isolated environment. If the RESTful API is dependent to other services, the later shall be replaced by mock components. We do not assume that those mock components exist, our approach builds them.
	
\end{itemize} 

\subsection{IOTS Test Case Definition}

Given the test architecture of Figure \ref{fig:archi}, we consider that events have the form $e(\alpha)$ with $e$ some label, e.g., a path or a status; "*" is a special notation representing any label. $\alpha$ is an assignment of parameters in $P$ to a value in the set of values $V$. These parameters allow the encoding of some specific web service characteristics e.g., if an event is a request, the receiver and sender of this request, etc. We write $x:=*$ the assignment of the parameter $x$ with an arbitrary element of $V$, which is not of interest. $\EuScript{E}$ denotes the event set. 
We also use these additional notations on an event $e(\alpha)$ to make our algorithm more readable: $from(e(\alpha)))$ (reps. $to(e(\alpha))$) denotes the source (resp. the destination) of the event. $isreq(e(\alpha))$, $isresp(e(\alpha))$ are boolean expressions expressing the nature of the event. $body(e(\alpha))$, $header(e(\alpha))$, $status(e(\alpha))$ are expressions returning values in $\alpha$.

We model a test case with a deterministic IOTS having a tree form and whose terminal states express test verdicts, e.g., $pass$ or $inc$, which stands for inconclusive. A test step corresponds to an IOTS transition $q \xrightarrow{e(\alpha),l} q'$ with $e(\alpha)$ some event and $l$ a label set, which may be empty. Furthermore, we use the notation $\theta$ labelled on transitions to represent the absence of reaction from a service under test \cite{Phillips1987RefusalT}. Classically, we call a sequence of test steps a test sequence. The label set allows to easily express some knowledge about the event. For instance, "crash" is used when the HTTP status 500 is received. The special label "mock" identifies events performed by some other dependee services. Since we assume testing $SUT$ in isolation, the dependee services will have to be replaced with mock components. 

An IOTS test case has to met a few restrictions to avoid nondeterministic behaviours while testing. To this end, a test case must allow at most one input event at any state. In reference to \cite{Tretmans2008}, this last restriction, we say that a test case is \textit{input restricted}. Additionally, still in the context of isolation testing and to keep control of the testing process, a mock component should be deterministic and return at most one response after being invoked with the same event. 
We say that a test case has to be mock response restricted. This is formulated with:

\begin{definition}
	\label{def:IOTS}
	A test case $tc$ is a deterministic IOTS $\langle Q,q0,\Sigma \cup \{ \theta \},L,\rightarrow \rangle$ where:
	\begin{itemize}
		\item $Q$ is a finite set of states; $q0$ is the initial state; 		
		
		\item $\Sigma \subset \EuScript{E}$ is the finite set of events. $\Sigma_I \subseteq \Sigma$ is the finite set of input events beginning with "?", $\Sigma_O \subseteq \Sigma$ is the finite set of output events beginning with "!", with  $\Sigma_O \cap \Sigma_I = \emptyset$;
		
		\item $L$ is a set of labels;
		
		\item $\rightarrow \subseteq Q \times \Sigma \cup \{\theta \} \times L^* \times Q$ is a finite set of transitions. A transition $(q,e(\alpha),l,q')$ is also denoted $q \xrightarrow{e(\alpha),l} q'$;
		
		\item $Q_f = \{pass, fail, inc\} \subset Q$ is the set of verdict states;
		if $q \xrightarrow{e(\alpha),l} q_f$ with $q_f\in Q_f$, then $e(\alpha) \in \Sigma_O \cup {\theta}$;
		
		\item $tc$ has no cycles except those in states of $Q_f$;
		
		\item $tc$ is input restricted i.e. $\forall q \in Q: 
		event(q) = \Sigma_O \cup \{e(\alpha)\}$ for some $e(\alpha) \in \Sigma_I$ or $event(q) = \Sigma_O\cup \{\theta\} $ with $event(q)=\{e(\alpha) \mid \exists q'\in Q: q \xrightarrow{e(\alpha),l}q'\}$;
		
		\item $tc$ is mock response restricted i.e. $\forall q \in Q:
		|\{q \xrightarrow{e(\alpha),l}q'\mid isResp(e(\alpha))\wedge mock \in l \}|\leq 1$.
	\end{itemize}
\end{definition}

\begin{figure}[htbp]
	\begin{center}
		\includegraphics[width=1\linewidth]{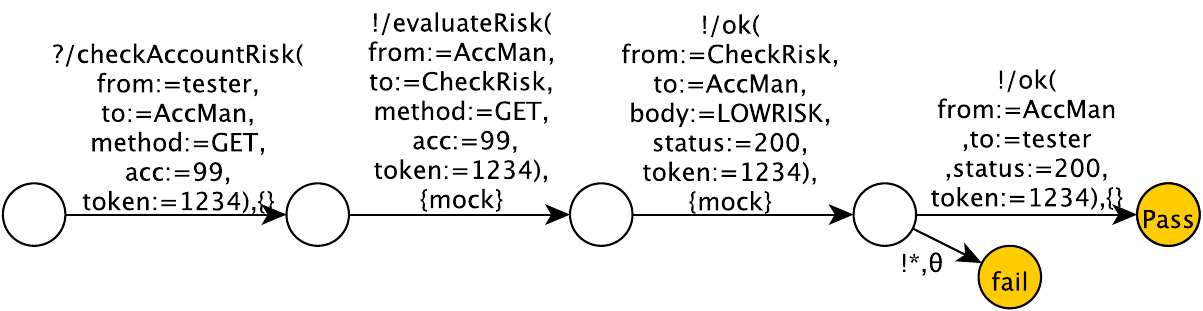}
	\end{center}
	\caption{IOTS Test Case example}
	\label{fig:tc0example}
\end{figure}


An IOTS test case example is illustrated in Figure \ref{fig:tc0example}. It checks whether a RESTful API AccMan can be called with "/checkAccountRisk". This service is dependent to another service called CheckRisk. The events related to CheckRisk are labelled by "mock" to express that a mock component has to be built to test AccMan in isolation.

IOTS test cases can be written manually, but this activity may be long and error-prone, especially for un-experimented developers. To solve this problem, we proposed in \cite{SalvaS23} an approach and tool for generating IOTS test cases from Log files. The approach also allows to recognise some specific behaviours (authentication, token generation, crash) and adds on test steps the following labels "login", "token", "token generation", "crash".

\subsection{Mutation Operators For Security Testing}

Based on our observations about the related work, we chose to define mutation operators specialised for the detection of weaknesses in RESTful APIs. We initially conducted a literature review to collect relevant data about the security testing of RESTFul APIs. 
We searched for papers indexed in online sources (Scopus, Science Direct, IEEE Xplore, ACM Digital Library, Google Scholar). We identified relevant papers via keyword search by using the terms ”web services security weaknesses vulnerability  attacks" and then, terms ”microservice security weaknesses vulnerabilities attacks”. We found 42 and 35 works between 2006-2023. We isolated 24 papers and 3 surveys by using their abstracts and titles. We then crossed  these results with the databases CAPEC \cite{CAPEC} and CWE \cite{CWE} of the MITRE organisation in order to classify attacks and avoid duplicates. With regard to our black box test architecture, we kept the attacks related to these domains:
\begin{itemize}
\item CAPEC-21: Exploitation of Trusted Identifiers
\item CAPEC-22: Exploiting Trust in Client
\item CAPEC-63: Cross-Site Scripting (XSS)
\item CAPEC-151: Identity Spoofing
\item CAPEC-153: Input Data Manipulation
\item CAPEC-115: Authentication Bypass
\item CAPEC-125: Flooding
\item CAPEC-278: Service Protocol Manipulation
\item CAPEC-594: Traffic Injection  
\end{itemize}

At this step, we collected a total number of 36 attacks. We finally augmented this compilation, by incorporating 7 recommendations provided in the ENISA good practice guide \cite{ENISA20}. Then, we studied these 43 elements to extract mutation operators. During this process, we applied the following criteria: 

\begin{itemize}
\item C1: in accordance with our test architecture, we build mutation operators applicable to unencrypted events;
\item C2: a mutation operator performs small changes, it is here used to build an attack executed with one test case only. Hence, complex attack scenarios cannot be considered;
\item C3: knowledge typically plays a crucial role in performing security attacks. We consider having labels in test steps allowing to recognise authentication processes, token generation and errors. Additional labels allow to recognise the existence of variables acting as tokens or session identifiers; 	
\item C4: an operator can derive new test cases and new mock components.
\end{itemize}
	
Using these criteria, we finally wrote 17 mutation operators tailored to testing in isolation the security of black box RESTful APIs. These operators are outlined in Table \ref{tab:mutationTable}, where column 2 provides the sources considered for constructing the operators, column 3 gives short descriptions, columns 4 and 5 give the expected behaviours that should be observed after the execution of mutated test steps and conditions on the application of the operators. 


\section{\uppercase{Test Case Mutation}}
\label{sec:approach}

%

	
\begin{figure}[htbp]
	\vspace{-0.3cm}
	\begin{center}
		\includegraphics[width=1\linewidth]{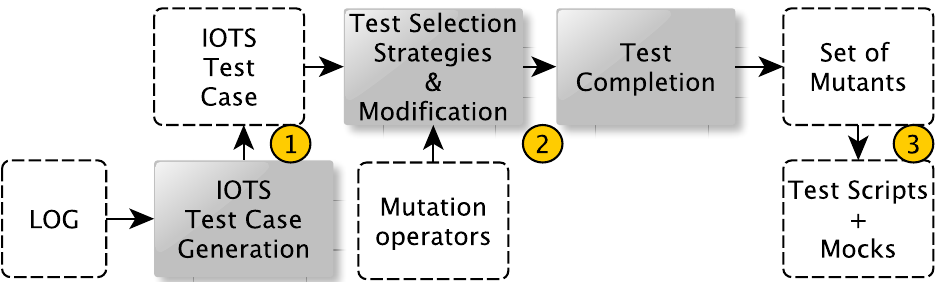}
	\end{center}
	\caption{Approach Overview}
	\label{fig:overview}
\end{figure}	
	
As illustrated in Figure \ref{fig:overview}, we propose a test case mutation approach and a tool for RESTful APIs, consisting of three main stages :

\begin{enumerate}

\item our approach takes either existing IOTS test cases,  or Log files that are used to generate IOTS test cases. As stated previously, the paper \cite{SalvaS23} presents algorithms and a tool for performing this step;

\item mutation operators are applied on IOTS test cases to perform slight modifications that aim to mimic security attacks. These modifications may result in numerous mutated test cases. To address this, we suggest strategies to restrict their generation. Mutation operators are then applied on test cases: we check whether the test steps meet some mutation conditions to restrict the transformations on the relevant steps only; we modify the original test cases  and complete them with tests steps and verdicts to get new IOTS mutants;

\item the mutants are finally converted into test scripts and mock components, which will be used to check whether $SUT$ is vulnerable.
\end{enumerate}

We formalise those steps in the remainder of this section. 
	
\subsection{Test Mutation Operator Definition}
\label{sec:mutation}


A mutation operator $M$ of an IOTS test case $tc$ is made up of three elements. The first is the function $Condition$, which aims at restricting the application of the operator to some events of $tc$. The next function $Change$ applies the mutation on $tc$ and produces an initial mutant $tc_m$. Finally, $Expected$ is a function that completes $tc_m$ with test sequences finished by  verdict states in order to express the expected observations after the execution of a mutated event.

\begin{definition}[Mutation operator]
A Mutation operator is the tuple $(Condition, Change, Expected)$ such that :

\begin{itemize}
\item $Condition : Q \times \Sigma \times L^* \times Q \rightarrow \{true, false\}$ is a function that expresses restrictions on test steps,

\item $Change : IOTS \rightarrow IOTS$ is a mutant derivation function,

\item $Expected : IOTS \rightarrow IOTS$ is a mutant completion function, such that for any test sequence  $q0 \xrightarrow{(e_1(\alpha_1),l_1)\dots (e_k(\alpha_k),l_k)} q$ of the IOTS, $q\in Q_f$ is a final state. 

\end{itemize}
\end{definition}

The function $Condition(q \xrightarrow{e(\alpha),l}q')$ of a mutation operator $M$ may be used on the label $e$, on the assignments $\alpha$, or on the label list $l$. This function can be used to define a generic operator, for example with a condition of the form $e==*$ supplemented with some conditions on $\alpha$. But, a more specific operator can also be defined with a condition on precise events and parameters. The last column of Table \ref{tab:mutationTable} provides several condition examples.  

$Change(tc)$ applies the mutation operator on a test case $tc$ and returns a mutant. We here assume having some transitions marked with the special label "mutation", which targets the transitions to transform. Given under the form of a procedure, $Change$ could have the following form :

\begin{lstlisting}[frame=single,escapeinside={(*}{*)}]
Reach a transition (*$t:= q\xrightarrow{e(\alpha),\{mutation\}}q_1$*);
Modify (*$t$*);
(possibly)Keep the next outgoing transitions from (*$q_1$*) to (*$q_k$*) such that (*$to(q_k\xrightarrow{e_k(\alpha_k)}q_l)=SUT$*);
Prune the useless transitions from (*$q_k$*) to a terminal state;
 \end{lstlisting}

$Expected(tc_m)$ completes a mutant returned by Change with new test steps such that the last test steps end by a verdict state. Column 4 of Table \ref{tab:mutationTable} summarises the test steps that are added for every mutation operator.


\subsection{Test Case Generation}

The test architecture of Figure \ref{fig:archi} emphasises the control and observation logics. The controller parts have the capability to send events to $SUT$. These events are those that can be modified by mutation operators to send unexpected requests or attacks. The observer parts will be used to collect responses, which are interpreted to decide whether $SUT$ is vulnerable or not.

In this context, we say that a test step $q \xrightarrow{e(\alpha),l}q'$ of a test case is mutable if the recipient of the event is $SUT$ itself and if the operator $M$ may be applied on this test step. Likewise, we use the notation $mutable(M) \text{ in } tc$ to get the set of test steps on which the mutation operator $M$ can be applied. It is worth noting that this set may be empty. This is captured by the following definition:

\begin{definition}[Mutable Test Step]
Let $M$ be a mutation operator, $tc$ be an IOTS test case for the service $SUT$, and $q \xrightarrow{e(\alpha),l}q' \in \rightarrow$ be a test step.

\begin{itemize}
	\item $q \xrightarrow{e(\alpha),l}q'$ is $mutable_M$ iff $to(e(\alpha))=SUT 
	\wedge M.Condition(q \xrightarrow{e(\alpha),l}q') 
	\wedge ((e(\alpha)\in \Sigma_I \vee "mock"\in l ))$.
	
	\item $mutable(M) \text{ in } tc=_{def} \{ q \xrightarrow{e(\alpha),l}q'\in tc \mid  q \xrightarrow{e(\alpha),l}q' \text{ is }mutable_M \}$
\end{itemize} 

\end{definition}

Furthermore, we define the IOTS operator $mark$, which simply adds a label "mutation" on the mutable test steps.

\begin{definition}[IOTS operator $mark$]
Let $t=q\xrightarrow{e(\alpha),l} q'$ be a test step of a test case $tc=\langle Q,q0,\Sigma \cup \{ \theta \},L,\rightarrow \rangle$.

$mark \text{ } t \text{ }in\text{ } tc = \langle Q_2,q0,\Sigma_2 \cup \{ \theta \},L_2,\rightarrow_2 \rangle$ is the IOTS test case derived from the test case $tc$ where $Q_2,\Sigma_2,L_2,\rightarrow_2$ are defined by the following rules:
\begin{center}
	\fbox{%
		\begin{minipage}{0.45\textwidth}
			\begin{center}
				\begin{tabular}{ll}
					$
					\frac{ t=q \xrightarrow{e(\alpha),l} q'} {q \xrightarrow{e(\alpha),l\cup \{"mutable"\}} q'}$&
					
					$\frac{t_2  \neq t} {t_2}
					$										
				\end{tabular}
			\end{center}
		\end{minipage}%
	}
\end{center}


%
%
\end{definition}


\begin{algorithm}[h]
\begin{scriptsize}
	\SetKwInOut{Input}{input} \SetKwInOut{Output}{output}
	
	\Input{Test case set $TC$, Mutation Operator $M$}
	\Output{Test case set $TC_M$}
	
	$TC_M:=\emptyset$\;
	\ForEach{$tc\in TC$}{
		\ForEach{$q \xrightarrow{e(\alpha),l}q' \in mutable(M) \text{ in } ts$ such that $ts = q0 \xrightarrow{(e_1(\alpha_1),l_1) \dots (e_k(\alpha_k),l_k)}pass \in tc$ and $selection(TC,TC_M)$}
		{
			$mark$ $q \xrightarrow{e(\alpha),l}q' \text{ }in\text{ } tc$ such that $q \xrightarrow{e(\alpha),l}q'$ $\in mutable(M) in \text{ } ts$ arbitrarly chosen\;
			$tc_2 :=M.Change(tc,q \xrightarrow{e(\alpha),l}q')$\;
			$tc_2 := M.Expected(tc_2)$\; 
			$compl$ $tc_2$\;
			$TC_M := TC_M \cup \{tc_2\}$\;
			
		}
		
}
			
\end{scriptsize}
\caption{IOTS Test Case Mutation}
\label{algo:TCGen}
\end{algorithm}



We are now ready to present our test case mutation algorithm given in Algorithm \ref{algo:TCGen}: it takes a mutation operator $M$ along with a test case set $TC$. It produces a new test case set, denoted $TC_M$. It covers every mutable test step of a test sequence $ts$ (line 3) starting from the initial state of the test case such that $ts$ is finished by the state $pass$. We choose to only mutate test sequences finished by $pass$ to avoid bringing confusion in the test result analysis. Indeed, if we mutate a test sequence finished by fail and if we obtain a fail verdict while testing, it is very difficult to deduce whether $SUT$ is faulty on account of the mutation. As the set of mutants may become large, Algorithm \ref{algo:TCGen} calls the function $selection(TC,TC_M)$, which returns a boolean value. This function expresses a mutant generation strategy, e.g., "applies $M$ on every test case only once", which stops the mutation of the test cases once some conditions are met. In this case, the function returns false. Algorithm \ref{algo:TCGen} marks the chosen test step with "mutable" to help the mutation operator target the test step to change. A new test case $tc_2$ is built by applying the function $M.Change$ and by completing its branches not finished by a verdict state with $M.Expected$ in order to express the expected behaviour after the execution of the mutated test step. Additionally, the mutant $tc_2$ is completed once more (line 7) with the operator $compl: IOTS \rightarrow IOTS$ to add transitions that express all the behaviours that might be observed and the related test verdicts. The resulting mutant $tc_2$ is stored in $TC_M$. The operator $compl$ is defined by:  

\begin{definition}[IOTS operator $compl$]
	$compl \text{ } tc = \langle Q_2,q0,\Sigma_2 \cup \{ \theta \},L,\rightarrow_2 \rangle$ is the IOTS test case obtained from $tc$ where $Q_2,\Sigma_2,\rightarrow_2$ are defined by the following rules: 
{\footnotesize 
	\setlength{\abovedisplayskip}{6pt}
	\setlength{\belowdisplayskip}{\abovedisplayskip}
	\setlength{\abovedisplayshortskip}{0pt}
	\setlength{\belowdisplayshortskip}{3pt}
\noindent \begin{align*}
r_1 :& q_1 \xrightarrow{e(\alpha),l} q_2 \vdash q_1 \xrightarrow{e(\alpha),l} q_{2}\\
r_2 :& q_1 \xrightarrow{e(\alpha),l} q_2, q_1 \xrightarrow{!*,\{\}} q_3 \notin \rightarrow \vdash q_1 \xrightarrow{!*,\{\}} inc\\
r_3 :& q_1 \xrightarrow{!e(\alpha),l} q_2, q_1 \xrightarrow{?e_2(\alpha_2),l}q_3 \notin \rightarrow, q_1 \xrightarrow{\theta}q_3 \notin \rightarrow \vdash\\
&q_1 \xrightarrow{\theta} fail
\end{align*}
}%
\end{definition}

The inference rule $r_1$ takes all the transitions of an IOTS to build a new test case. $r_2$ completes the test case with a new transition to express that any unexpected output leads to the inconclusive verdict. When the test case only expects outgoing transitions labelled by output events, the rule $r_3$ also adds a transition to fail modelling that the absence of reaction is faulty.


The function $selection(TC,TC_M)$ encodes conditions on the test case sets $TC$ and $TC_M$ to limit the number of mutants by mutation operator. Various conditions and combinations could be considered. Here, we provide some examples:

\begin{itemize}
\item No restriction (all mutable test steps are covered);
\item Every test case is mutated at most $n$ times
\item Every mutable test step of each test case is mutated at most $n$ times

\end{itemize}
The impact of these strategies will be studied in Section \ref{sec:evaluation}.

\begin{figure}[htbp]
	\begin{center}
		\includegraphics[width=1\linewidth]{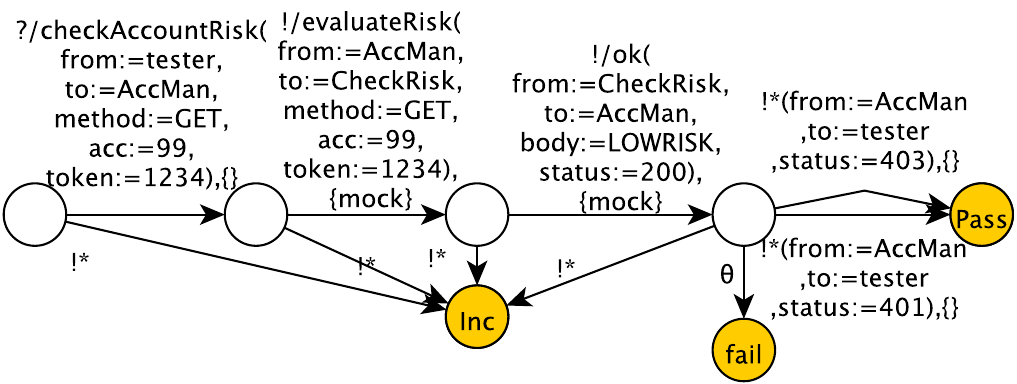}
	\end{center}
	\caption{IOTS Mutated Test Case example}
	\label{fig:tc0m}
\end{figure}

Figure \ref{fig:tc0m} illustrates an example of mutant obtained from the test case of Figure \ref{fig:tc0example} by applying the operator "Token removal" on the second mutable test step (!/ok), which is performed by a mock component. The verdict is pass if a response is observed with an HTTP status 401 or 403, which encode that the request has been rejected on account of insufficient permissions. 



\subsection{Generation of Concrete Test Cases}
\label{sec:TCGenComponent}

Finally, executable test scripts are generated from IOTS test cases. 
We have chosen to generate test cases using the frameworks Citrus and Mockserver. Given an IOTS test case $tc\in TC_M$, some parameters may still be assigned to "*". For example, this happens for parameters used to identify sessions. In short, we assign these parameters with stored values collected from Log files or with random values. To generate a test script, the transitions of $tc$ labelled by "mock" are initially pruned. The resulting IOTS is converted as follows: every request to $SUT$ is converted into code that calls $SUT$ and waits for a response. An example is given in Figure \ref{fig:concretetestcase}. The transitions labelled by responses of this request are used to build assertions. 
The test script ends with the call of the method "verificationMock", which aims to check whether mock components behave as expected during the test execution. At the moment, we check whether the number of calls to a mocked request matches with the number of time this request is found in $tc$. 

To generate mock components, the IOTS transitions of $tc$ labelled by "mock" are used to derive rules of the form \textit{request()...respond()}, which mimic the behaviour of a dependee service. Then, the method "verificationMock" is written according to these rules. Figure \ref{fig:concretetestcase2} shows a rule example written with the language provided by the framework MockServer. 

\begin{figure}[htbp]		
	\vspace{-0.3cm}
\begin{lstlisting}[frame=single, numbers=left,  numberstyle=\tiny, stepnumber=2]
@Test @CitrusTest
public void testAccMan() throws FileNotFoundException{
HttpClient toClient = CitrusEndpoints
.http().client().requestUrl("http://AccMan/").build();
$(HTTP()
.client(toClient).send().get("checkAccountRisk").message()
.header("token",1234).body("\"acc\"=99")
.accept(MediaType.ALL_VALUE));
$(receive(toClient)
.message().type(MessageType.PLAINTEXT).name("Response")
.extract(fromHeaders()
.header(HttpMessageHeaders.HTTP_STATUS_CODE, "statusCode"))
.header("token","token")));
variable("body","citrus:message(Response.body())");
variable("status", "${statusCode}");
String status = context.getVariable("status");
String t = context.getVariable("token");
If (token.equals("1234") && status.equals("403")) assertTrue(true);
else Assumptions.assumeTrue(false,"Inconclusive");
verificationMock();}
\end{lstlisting}
\caption{Example of test script for the service AccMan}
\label{fig:concretetestcase}
\end{figure}

\begin{figure}[htbp]		
\vspace{-0.3cm}
\begin{lstlisting}[frame=single,numbers=left,numberstyle=\tiny,stepnumber=2]
mockServer.when(
request().withMethod("GET").withPath("/evaluateRisk")
.withHeaders( new Header("acc", "99"),new Header("token", "1234"))
,Times.exactly(1))
.respond( response().withStatusCode(200)
.withBody("LOWRISK"));    
	\end{lstlisting}
	\caption{Mock component piece of code, which implements the events !/EvaluateRisk and !ok of the test case of Figure \ref{fig:tc0m}}
	\label{fig:concretetestcase2}
\end{figure}

\section{\uppercase{Preliminary Evaluation}}
\label{sec:evaluation}

This evaluation aims at investigating the capabilities of our algorithm through the following questions:

\begin{itemize}
	\item RQ1: how many mutants are generated ? Do our strategies succeed at reducing the number of generated mutants? 
	\item RQ2: are the mutants effective to uncover weaknesses or to increase code coverage?
	\item RQ3: what is the performance of our algorithm?
\end{itemize}

This study was conducted on four case studies:

\begin{itemize}
	\item C1: Piggy metric\footnote{\url{https://github.com/sqshq/piggymetrics}} is a financial advisor application composed of 3 micro-services specialised in account management, statistics generation and notification management; 
	\item C2: eShopOnContainers\footnote{\url{https://github.com/dotnet/eShop}},  implementing an e-commerce web site using a services-based architecture (5 RESTful APIs);
	\item C3: a loan approval process implemented with 4 RESTful APIs developed by third year computer science undergraduate students; 
	\item C4: a composition of 3 RESTful APIs used to implement an online shop (stock management, client management, purchase, etc.) still developed by students.
	
\end{itemize}

Log files were collected by applying scenarios performed manually and by executing the penetration testing tool ZAProxy. We collected 16603 HTTP messages for C1, 76220 for C2, and 10000 for C3 and for C4.

We implemented Algorithm \ref{algo:TCGen} in a prototype tool, which takes IOTS test cases stored in Json files or Log files directly and generates mutants along with executable test scripts and mock components. This tool integrates 4 mutation operators, namely HTTP Verb Change, Path Manipulation, Session Management and Token Removal, which refer to severe and frequent vulnerabilities. It also implements 3 test case generation strategies: S0 that does not restrict the number of mutants, S1 that produces at most one mutant for every event that belongs to a mutable test step, and S2 that returns at most 2 mutants per test case in $TC$. In comparison to S0, S1 is limiting when the initial test case uses the same event several times. The tool, source codes and Log files are available in \cite{companion}.

\subsection{RQ1: how many mutants are generated?}
To answer this question, we measured the number of mutants generated with each strategy for every RESTful API across the four case studies, while varying the size of the initial test case set $TC$. The initial test case sizes were between 6 to 10 test steps.

\begin{figure}[htbp]
	\begin{center}
		\includegraphics[width=1\linewidth]{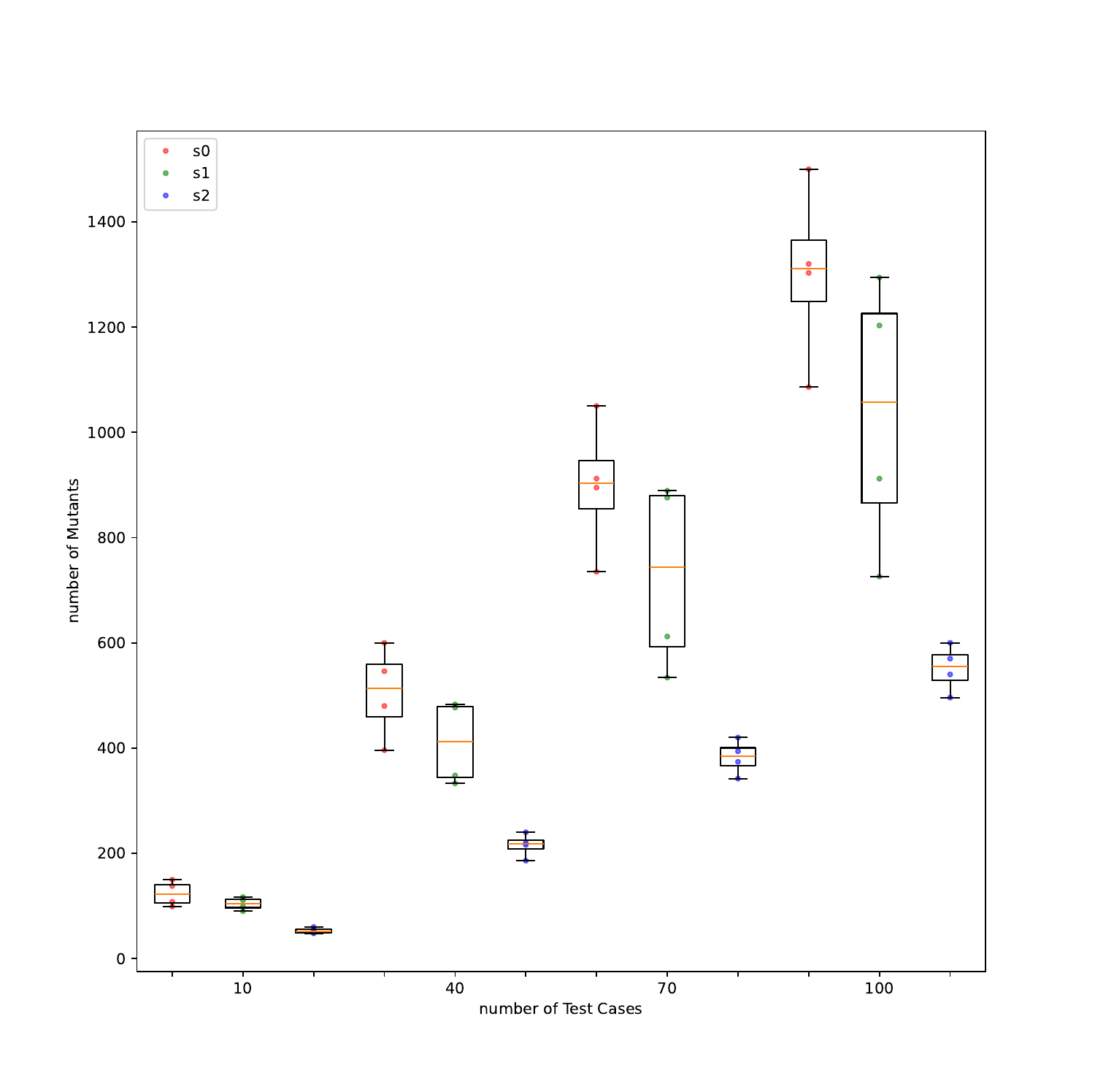}
	\end{center}
	\vspace{-0.1cm}
	\caption{\# mutants vs. \# test cases}
	\label{fig:tc_nb}
\end{figure}

Figure \ref {fig:tc_nb} summarises our measures with a boxplot, which shows the number of mutants generated for S0, S1 and S2 with 10, 40, 70 and 100 original test cases. 
We observed that the number of mutants follows a linear relationship, which means that the mutant set sizes are consistent with the sizes of the initial test case sets.  The boxplot shows significant differences among the strategies. In average, we have a test case ratio increase ($\frac{\text{\# mutants} - \text{\# initial test cases}}{\text{\# initial test cases}}$) of 1180\% for S0, 940\% for S1, 440\% for S2. In comparison to S0 (generation of all the possible mutants), S1 reduces the number of mutants at most with a ratio of 1:1.5, but sometimes the number of mutants generated with S1 is close to the number obtained with S0. This is explained by the fact that S1 depends on the occurrences of the events in a test case, which may vary. With S2, the number of mutants at most reduced with a ratio of 1:2.5. 
Both S1 and S2 appear to be interesting for limiting the mutant generation, on condition to keep good effectiveness, which is studied in the next section.



\subsection{RQ2: Are the mutants effective?}

To investigate this question, we experimented our algorithm on C3 and C4. As these service compositions were coded by students, we expect the detection of security defects. We reviewed the source code to list the weaknesses targeted by our 4 mutation operators and deduced for C3 that all the services have two weaknesses (access restrictions bypass with HTTP verbs; insufficient session expiration) and for C4 that all the services also include two weaknesses (unauthorized access with token removal; insufficient session expiration), which can be repeated several times. For each strategy and service, we measured the number of mutants, the ratio of failed tests while executing these mutants, the number of weaknesses detected by the new test cases and the increase of coverage between the RESTful APIs testing without and with mutants (coverage with $TC+TC_M$ - code coverage with $TC$ only).

\begin{table*}[]
	\caption{Mutant effectiveness evaluation for C3 and C4: col.3:number of original test cases, col.4: number of mutants, col.5: \% of failed mutants, col.6: number of weaknesses detected by the mutants / number of observed weaknesses, and col.7,8: line and branch coverage increase}
	\label{table:rq2}
	\scriptsize
		\begin{tabular}{|l|l|l|l|l|l|l|l|}
			\hline
		RESTful API&Strategy&$|TC|$&$|TC\_M|$&\% failed tests& \# vul. / \# obs vul. &Line coverage increase&Branch coverage increase\\ \hline
		\multirow[t]{3}{*}{C4:Cust.Inter.}& S0&8&26&65\%&2 / 2&7\%&15\%\\ 
		& S1&8&26&65\%&2 / 2&7\%&15\%\\ 
		& S2&8&11&36\%&1 / 2&12\%&3\%\\ \hline
		\multirow[t]{3}{*}{C4:PaymentAndCards} & S0&6&21&43\%&2 / 2&3\%&6\%\\
		& S1&6&20&40\%&2 / 2&3\%&6\%\\ 
		& S2&6&12&66\%&2 / 2&2\%&4\%\\	\hline
		
		\multirow[t]{3}{*}{C4:Prod.Manag.} 	& S0&11&44&52\%&2 / 2&7\%&16\%\\
		& S1&11&36&58\%&2 / 2&7\%&16\%\\ 
		& S2&11&11&54\%&2 / 2&3\%&6\%\\ \hline
		\multirow[t]{3}{*}{C3:Acc.Manag.} 	& S0&12&96&99\%&5 / 5&24\%&17\%\\
		& S1&12&71&99\%&5 / 5&24\%&17\%\\ 
		& S2&12&12&92\%&5 / 5&24\%&17\%\\ \hline	
		\multirow[t]{3}{*}{C3:App.Manag.} 	& S0&8&35&94\%&5 / 5&33\%&0\%\\
		& S1&8&34&99\%&5 / 5&33\%&0\%\\ 
		& S2&8&12&92\%&5 / 5&33\%&0\%\\	\hline
		\multirow[t]{3}{*}{C3:Loan.App} 	& S0&4&47&89\%&4 / 4&32\%&0\%\\
		& S1&4&42&81\%&4 / 4&32\%&0\%\\ 
		& S2&4&12&50\%&4 / 4&28\%&0\%\\	\hline
		\multirow[t]{3}{*}{C3:Check.acc} 	& S0&6&35&86\%&4 / 4&23\%&25\%\\
		& S1&6&51&81\%&4 / 4&23\%&25\%\\ 
		& S2&6&12&50\%&4 / 4&23\%&25\%\\ \hline               
	\end{tabular}
\end{table*}


Table \ref{table:rq2} shows the results by RESTful API. We can deduce that all the strategies allow to detect weaknesses, but the mutants obtained by both S0 and S1 are more effective than those produced by S2 as one weakness is not detected with S2 (C4:Cust.Inter.). In line with this result, the branch coverage is increased in average on all the services by 18,4\% with S1 and S2, and by 17,8\% with S2. With regard to line coverage, we observe an increase of 13,1\%, 12,1\% and 9,1\% for S0, S1, S2. These results suggest that the mutants generated with the three strategies are effective to uncover defects but both S0 and S1 are slightly more effective.


\subsection{RQ3: what is the performance of our algorithm?}

To answer RQ3, we studied the two factors that influence the complexity of Algorithm \ref{algo:TCGen}, i.e. the size of the initial test case set $TC$ and the length of these test cases. These experiments were carried out on a computer with 1 Intel CPU i7-4790 @ 3.6GHz and 16GB RAM. For the first factor, we arbitrarily chose to consider C3. We built initial sets $TC$ by varying the number of test cases between 10 to 100 having around 10 events. For the second factor, we took back C3 and built sets of 20 test cases, by varying their length from 4 to 300 events. Then, we measured execution times. 

\begin{figure}[htbp]
	\begin{center}
		\includegraphics[width=1\linewidth]{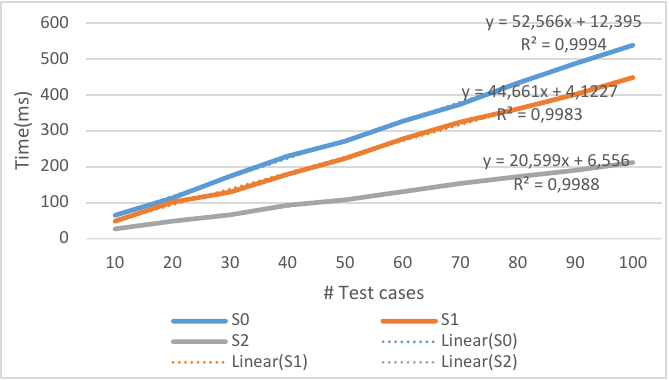}
	\end{center}
	\caption{Execution time vs. \# test cases}
	\label{fig:timenbtc}
\end{figure}

Figure \ref{fig:timenbtc} depicts execution times in milliseconds  w.r.t. the size of the initial test case set $TC$. We observe that our algorithm ends quickly (less than 1s) even with large test case sets. The curves follow a linear regression, showing that Algorithm \ref{algo:TCGen} scales well with the size of $TC$. Unsurprisingly, the test case strategy S2, which limits the generation of 2 mutant by original test case, achieves faster computation. 

\begin{figure}[htbp]
	\begin{center}
		\includegraphics[width=1\linewidth]{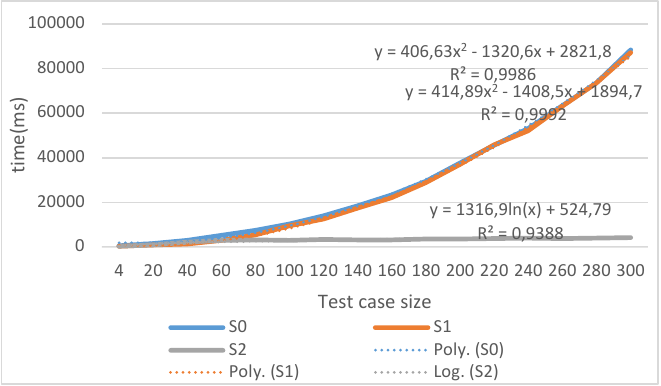}
	\end{center}
	\caption{Execution time vs. test case size}
	\label{fig:timesizetc}
\end{figure}

Figure \ref{fig:timesizetc} also depicts execution times in seconds w.r.t. the length of the test cases. At worst (20 test cases of 300 events), 25 minutes are required to generate mutants with the strategies S0 or S1. For these, the curves follow a quadratic regression, expressing that Algorithm \ref{algo:TCGen} does not scale well with the test case size. But in practice, it is uncommon to have test cases with more than 20 steps. As previously, the strategy S2 exposes quicker executions as only 2 mutants are built by operator, independently on the test case size.


In summary, these experiments tend to show that our algorithm can be used in practice to generate new security test cases, which are effective to detect new defects or to better cover the service codes. The strategies S0 and S1 produce slightly more effective test cases than S2 but are less efficient, especially if the test case length is high. In view of the three studied strategies, S1 seems to be the most appropriate if effectiveness is a priority.



\section{\uppercase{Conclusion}}
\label{sec:conclusion}

In this paper, we present an original solution to generate mutated test cases for testing the security of RESTful APIs. We proposed a list of 17 mutation operators specialised in the generation of security test cases. Subsequently, we introduced an algorithm allowing to generate concrete test scripts and mock components by means of these operators. A significant contribution of this algorithm is its ability to generate mock components to test a RESTful API in isolation. We evaluated this algorithm through four case studies. Our results demonstrate its capability to construct hundreds of test cases and mock components within minutes, and show good scalability. Besides, the mutants enable the detection of weaknesses and enhance code coverage.
 
At the moment, our mutation operators allow to infer mutants that mimic attacks performed by one test step. As part of future work, we aim to define more sophisticated operators that could support the mutation of several steps at a time, thus constructing more complex attack scenarios.





\bibliographystyle{apalike}
{\small \bibliography{doc} }

\noindent 
\afterpage{\onecolumn
	\scriptsize
	\begin{center}
		\topcaption{Test case mutation operators for the security testing of RESTful APIs} 
		\tablefirsthead{}
		\tablehead{}
		\tabletail{\hline}
		\tablelasttail{}
		\label{tab:mutationTable}
\begin{xtabular*}{\textwidth}{|p{1.3cm}|p{2.3cm}|p{2.3cm}|p{5.3cm}|p{2.5cm}|}
\hline
\textbf{Mutation} & \textbf{source} & \textbf{Description}&\textbf{Expected Behaviour}&\textbf{Mutation Condition}\\
		\hhline{|=|=|=|=|=|}
		Event Duplication&\cite{ENISA20} TM-20 &duplicate a request event to $SUT$&
		$q \xrightarrow{!*(\alpha),l} pass$ with $"crash"\notin l \wedge from(q \xrightarrow{!*(\alpha),l} pass)=SUT$
		&$e==*$\\
		\hline
		HTTP Verb Change&CAPEC-274&changing the HTTP verb of a request 
		&
		$q \xrightarrow{!*(\alpha),l} pass$ with $status(\alpha):=405 \wedge from(q \xrightarrow{!*(\alpha),l} pass)=SUT$
		&isReq($e(\alpha)$)==true\\
		\hline
		XSS attack &\cite{ENISA20}
		TM-21, CAPEC-63,
		CWE-79, \cite{SecuringvulnerablehomeIoTdeviceswithanin-hubsecuritymanager},\cite{OnaClassificationApproachforSOAVulnerabilities}&XSS attack&
		$q \xrightarrow{!*(\alpha),l} pass$ with $"crash"\notin l \wedge contains("error", body(\alpha)) \wedge from(q \xrightarrow{!*(\alpha),l} pass)=SUT$ 
		&$body(\alpha)!="" \vee header(\alpha)!=""$\\
		\hline
		Cryptographic failures&CAPEC-220, \cite{ENISA20} PS-15, \cite{OntheNatureofIssuesinFiveOpenSourceMicroservicesSystems:AnEmpiricalStudy}, CAPEC-276, \cite{SecuringvulnerablehomeIoTdeviceswithanin-hubsecuritymanager}, \cite{OnaClassificationApproachforSOAVulnerabilities}, CWE-287&
		Replay an event using untrusted connexion&
		$q \xrightarrow{!*(\alpha),l} pass$ with $contains("ERR\_CERT\_AUTHORITY\_INVALID",$ $body(\alpha)) \wedge from(q \xrightarrow{!*(\alpha),l} pass)=SUT$ 
		&$e=="*"$\\
		\hline
		Token Removal&CWE-602, CWE-862, \cite{OnaClassificationApproachforSOAVulnerabilities}, CAPEC-114&
		delete a token in event&
		$q \xrightarrow{!*(\alpha),l} pass$ with $status(\alpha)= 401 \vee 
		status(\alpha)= 403\wedge from(q \xrightarrow{!*(\alpha),l} pass)=SUT$
		&"token" $\in$ l\\
		\hline
		Token Removal on the creation&CWE-602, CWE-862, \cite{OnaClassificationApproachforSOAVulnerabilities}, CAPEC-114&
		delete a token in event&
		$q \xrightarrow{!*(\alpha),l} pass$ with $status(\alpha)\geq 401 \wedge  
		status(\alpha)\leq 403 \wedge from(q \xrightarrow{!*(\alpha),l} pass)=SUT$&
		"token creation" $\in$ l\\
		\hline
		Token Alteration&\cite{OnaClassificationApproachforSOAVulnerabilities},\cite{OntheNatureofIssuesinFiveOpenSourceMicroservicesSystems:AnEmpiricalStudy}, CAPEC-114&replacing a token of an event by another one if possible. three types: expired authentication token, token existing but not for this session, and token not existing&
		$q \xrightarrow{!*(\alpha),l} pass$ with $status(\alpha)\geq 401 \wedge  
		status(\alpha)\leq 403 \wedge from(q \xrightarrow{!*(\alpha),l} pass)=SUT$&
		"token" $\in l$\\
		\hline
		
		Stress Testing &\cite{SoK:Run-timesecurityforcloudmicroservices.Arewethereyet}, \cite{CybersecurityforserviceorientedarchitecturesinaWeb2.0world:AnoverviewofSOAvulnerabilitiesinfinancialservices}, CAPEC-488&replay events to the tested service a lot of times in a small window&
		$q \xrightarrow{!*(\alpha),l} pass$ with $"crash"\notin l \wedge \neg contains("error" body(\alpha)) \wedge from(q \xrightarrow{!*(\alpha),l} pass)=SUT$&
		$e=="*"$\\
		\hline 
		SSRF Enforce “deny by default” &\cite{ENISA20}&request or response from an unknown service&
		$q \xrightarrow{!*(\alpha),l} pass$ with $"crash"\notin l \wedge (contains("error",body(\alpha)) \vee status(\alpha):=404) \wedge from(q \xrightarrow{!*(\alpha),l} pass)=SUT$&
		$e=="*"$\\
		\hline
		
		Body data manipulation&\cite{ENISA20} TM-06, \cite{IntegrityProtectionAgainstInsidersinMicroservice-BasedInfrastructures:FromThreatstoaSecurityFramework}, CAPEC-278, CAPEC-92, CWE-20, CWE-125&replay request using unauthorized data&
		$q \xrightarrow{!*(\alpha),l} pass$ with $(status(\alpha):=400 \vee status(\alpha):=422) \wedge from(q \xrightarrow{!*(\alpha),l} pass)=SUT$&
		$body(\alpha)!="" \vee header(\alpha)!=""$
		\\
		\hline 
		
		Cookie manipulation&CWE-472, CAPEC-31, \cite{OntheNatureofIssuesinFiveOpenSourceMicroservicesSystems:AnEmpiricalStudy}& change a cookie to inject an attack&
		$q \xrightarrow{!*(\alpha),l} pass$ with $status(\alpha):=400 \wedge from(q \xrightarrow{!*(\alpha),l} pass)=SUT$&		
		$cookies(\alpha)!=""$\\
		\hline
		Failed Login Attempt Duplication&\cite{ENISA20} TM-38, CAPEC-49&duplicating login event with wrong credentials&
		$q \xrightarrow{!*(\alpha),l} pass$ with $"crash"\notin l \wedge contains("error : Too Many Failed Attempt", body(\alpha)) \wedge from(q \xrightarrow{!*(\alpha),l} pass)=SUT$&
		$isReq(e(\alpha)) \wedge "login"\in l$\\
		\hline
\end{xtabular*}
\end{center}
\twocolumn}
  \newpage
\noindent 
\afterpage{\onecolumn
	\scriptsize
	\begin{center}
		\topcaption{Test case mutation operators for the security testing of RESTful APIs} 
		\tablefirsthead{}
		\tablehead{}
		\tabletail{\hline}
		\tablelasttail{}
		\label{tab:mutationTable2}
\begin{xtabular*}{\textwidth}{|p{1.3cm}|p{2.3cm}|p{2.3cm}|p{5.3cm}|p{2.5cm}|}
\hline
\textbf{Mutation} & \textbf{source} & \textbf{Description}&\textbf{Expected Behaviour}&\textbf{Mutation Condition}\\
		\hhline{|=|=|=|=|=|}
		Path manipulation &CWE-22,CAPEC-126, \cite{OnaClassificationApproachforSOAVulnerabilities}&change URL to get unauthorised access to data&
		$q \xrightarrow{!*(\alpha),l} pass$ with $status(\alpha):=404 \wedge from(q \xrightarrow{!*(\alpha),l} pass)=SUT$&
		$isReq(e(\alpha))$\\
		\hline
		SQL injection&CWE-89, \cite{OnaClassificationApproachforSOAVulnerabilities}, CAPEC-66 & manipulate input data to inject SQL code&
		$q \xrightarrow{!*(\alpha),l} pass$ with $(status(\alpha):=400 \vee contains("error", body(\alpha))) \wedge from(q \xrightarrow{!*(\alpha),l} pass)=SUT$&
		$body(\alpha) !=""$\\
		\hline 
		Session management &\cite{CybersecurityforserviceorientedarchitecturesinaWeb2.0world:AnoverviewofSOAvulnerabilitiesinfinancialservices}, CAPEC-61, CWE-613&add a (long) delay during which no reaction should be observed before the next event&
		$q \xrightarrow{!*(\alpha),l} pass$ with $status(\alpha):=401 \wedge contains("error: session terminated" body(\alpha))\wedge from(q \xrightarrow{!*(\alpha),l} pass)=SUT$&
		$e=="*"$\\
		\hline
		Information leakage&\cite{CybersecurityforserviceorientedarchitecturesinaWeb2.0world:AnoverviewofSOAvulnerabilitiesinfinancialservices}, CWE-200, \cite{IntegrityProtectionAgainstInsidersinMicroservice-BasedInfrastructures:FromThreatstoaSecurityFramework}&modify a request to get access to sensitive information&
		$q \xrightarrow{!*(\alpha),l} pass$ with $status(\alpha):=401 \wedge "crash"\notin l \wedge from(q \xrightarrow{!*(\alpha),l} pass)=SUT$&
		$isReq(e(\alpha))$\\
		\hline
		Dependee service shutdown&&shutdown a mock component after requesting it&
		$q \xrightarrow{!*(\alpha),l} pass$ with $"crash"\notin l \wedge (contains("error : connexion timed out", body(\alpha)) \vee status(\alpha):=408) \wedge from(q \xrightarrow{!*(\alpha),l} pass)=SUT$&		
		$q \xrightarrow{*(\alpha),l}$ with $"mock"\in l$\\
		\hline
		
		Buffer overflow &Capec-100, CWE-119 &overflow input data for trying to crash a server&
		$q \xrightarrow{!*(\alpha),l} pass$ with $status(\alpha):=400 \wedge "crash \notin l \wedge from(q \xrightarrow{!*(\alpha),l} pass)=SUT$&
		$e=="*"$\\
		\hline
		\end{xtabular*}
	\end{center}
\twocolumn}

\end{document}